\documentclass[prb,twocolumn,showpacs,preprintnumbers,amsmath,amssymb,superscriptaddress]{revtex4}
\usepackage{graphicx}% Include figure files
\usepackage{dcolumn}% Align table columns on decimal point
\usepackage{bm}% bold math

\def\Vec#1{\bm{#1}}
\begin{document}

%\preprint{}

\title{
Determination of the pairing state in iron-based superconductors through neutron scattering
}

\author{Yuki Nagai}
\affiliation{CCSE, Japan  Atomic Energy Agency, 5-1-5 Kashiwanoha, Kashiwa, Chiba, 277-8587, Japan}
\affiliation{CREST(JST), 4-1-8 Honcho, Kawaguchi, Saitama, 332-0012, Japan}
\affiliation{TRIP(JST), Chiyoda, Tokyo 102-0075, Japan}
\author{Kazuhiko Kuroki}
\affiliation{Department of Applied Physics and Chemistry, The University of Electro-Communications, Chofu, Tokyo 182-8585, Japan}
\affiliation{TRIP(JST), Chiyoda, Tokyo 102-0075, Japan}

%$^{2}$CREST(JST), 4-1-8 Honcho, Kawaguchi, Saitama, 332-0012, Japan\\
%$^{3}$TRIP(JST), Chiyoda, Tokyo, 102-0075, Japan
%}

\date{\today}% It is always \today, today,

\begin{abstract}
We calculate the spin susceptibility in the $s_\pm$ and $s_{++}$ 
superconducting states  
of the iron pnictides using the effective five orbital model 
and considering the quasiparticle damping.
For the experimentally evaluated magnitude of the 
quasiparticle damping and the 
superconducting gap,  the results at the wave vector $\simeq (\pi,0)$ 
show that the $s_\pm$ state is more consistent with the 
neutron scattering experiments, while 
for larger quasiparticle damping and the 
superconducting gap, the $s_{++}$ state can be more consistent.
To distinguish between two cases that reproduce the experiments at the 
wave vector $\simeq (\pi,0)$, we propose to investigate 
experimentally the wave vector $\simeq (\pi,\pi)$. 
\end{abstract}

\maketitle
%%%%%%%%%

%%%%%%%%%
The discovery of the iron-based superconductors has received 
considerable attention.  
The high transition temperature $T_{c}$\cite{Kamihara} itself 
has given a dramatic impact, but the possibility 
of peculiar unconventional pairing state has also been an issue of 
great interest. In fact, it has been proposed theoretically 
at the early stage
\cite{BangPRB2008,SeoPRL2008,ParishPRB2008,KorshunovPRB2008,Mazin,KurokiPRL}, 
that a natural pairing state 
is the so called $s_{\pm}$-wave pairing, where the superconducting gap 
is fully open, 
but changes its sign across the wave vector that bridges the 
disconnected Fermi surfaces. The sign change occurs because the spin 
fluctuations that develop at the Fermi surface nesting 
vector gives repulsive pairing interaction.  

There are many experimental results which suggest 
that the order parameter  
is fully gapped in a number of iron-based materials, 
such as the penetration depth\cite{Hashimoto,LuetkensPRL2008,MaronePRB2009} 
and the angular-resolved photo-emission 
spectroscopy (ARPES).\cite{DingEPL,Nakamura,Evtushinsky} 
There are also experimental suggestions 
that the iron-based materials are unconventional superconductors. 
The nuclear magnetic relaxation rate lacks coherence 
peak below $T_{c}$\cite{Nakai,Grafe,MukudaJPSJ08,TerasakiFeNMR}, 
which suggests the presence of 
sign change in the superconducting gap. 
%--------------2011/4/21 nagai-------------------start
%
Also in ref.~\cite{IBM}, integer and half-integer flux-quantum
transitions in composite niobium-ion pnictide loops
have been observed, 
which suggests the presence of the sign
change in the superconducting gap. In ref.~\cite{HanaguriFe},
STM/STS measurements have been performed to
detect the quasiparticle interference indicating the
realization of the $s_{\pm}$ gap.
%
%--------------2011/4/21 nagai-------------------end
%Also \cite{IBM}.  
These experiments seem to be consistent with the $s_{\pm}$ scenario.
However, recently there has been some debate concerning the 
sensitivity of $T_c$ against impurities. 
Although some experiments show suppression of $T_c$ by impurities\cite{Guo},
it has been pointed out, e.g., in refs.~\cite{Sato,Li} that 
the suppression of $T_c$ by impurities is too weak for 
a pairing state with a sign change in the gap.
A calculation based on a five band model by Onari {\it et al.} 
has supported this theoretically\cite{OnariKontani}, 
although the strength of the 
impurity potential adopted there is large compared to 
those calculated from first principles\cite{NakamuraArita}.
As a possible pairing state that is robust against impurities, 
the so-called $s_{++}$ state, where the gap does not change its sign 
between the Fermi surfaces, has been proposed\cite{Ono,Kontani,Saito}.

Now, it has been proposed at the early
stage\cite{Maier2,Maier,KorshunovPRB2008} 
that one of the promising ways to determine whether the 
superconducting gap indeed changes its sign between the disconnected 
Fermi surfaces is the observation of neutron scattering resonance
at the nesting vector of the electron and hole Fermi surfaces.
In fact, neutron scattering experiments have indeed 
observed a peak like structure in the 
superconducting state\cite{Christianson,Qiu,Inosov,Zhao,Ishikado}. 
This has been taken as a strong evidence for the sign change 
in the superconducting gap.
However, Onari {\it et al.} later took into account the quasiparticle damping 
effect in the calculation of the dynamical spin susceptibility, 
and showed that a peak like enhancement over the normal state values 
can be seen even in the $s_{++}$ state, which is 
due to the suppression of the 
normal state susceptibility originating from the damping\cite{Onari}.
In ref.~\cite{Onari}, the strength of the quasiparticle damping 
was estimated from the 
experimental results to be $\sim 10$(meV) or less, 
but larger values ($>50$(meV)) was adopted in the actual
calculation, fixing the gap/damping ratio to be around unity,  
due to the restriction in the numerical calculation.

In the present study, we revisit the problem of the resonance peak in
the neutron scattering experiment. We adopt the same formalism as in 
ref.~\cite{Onari}, but with smaller and realistic values ($\simeq 10$(meV)) of 
the quasiparticle damping and the superconducting gap by taking 
$k$-point meshes up to 16384$\times$ 16384. 
For such small values of the quasiparticle damping and the 
superconducting gap, 
the resonance peak enhancement over the normal state susceptibility 
at the wave vector $\simeq (\pi,0)$ is  
found to be comparable to those observed experimentally, while  
the enhancement over the 
normal state susceptibility for the $s_{++}$ state remains to
be small.
As we increase the quasiparticle damping and the superconducting gap 
up to values of an order of magnitude larger, 
the enhancement over the normal state susceptibility comparable to the 
experimental results arises in the $s_{++}$ state, 
in agreement with the results in ref.\cite{Onari}. 
For such a large quasiparticle damping and superconducting gap, 
an enhancement also occurs at the wave vector $(\pi,\pi)$. 
This is in contrast to the case with the small 
damping, where such an enhancement does not occur at $(\pi,\pi)$ (regardless
of the pairing state). Thus, we propose that by looking also at the 
wave vector $(\pi,\pi)$ in the neutron scattering experiments, 
one can distinguish the origin of the ``resonance like'' enhancement at 
$(\pi,0)$.

\begin{figure}[t]
  \begin{center}
    \begin{tabular}{p{65mm}}%  p{28mm}}
      \resizebox{65mm}{!}{\includegraphics{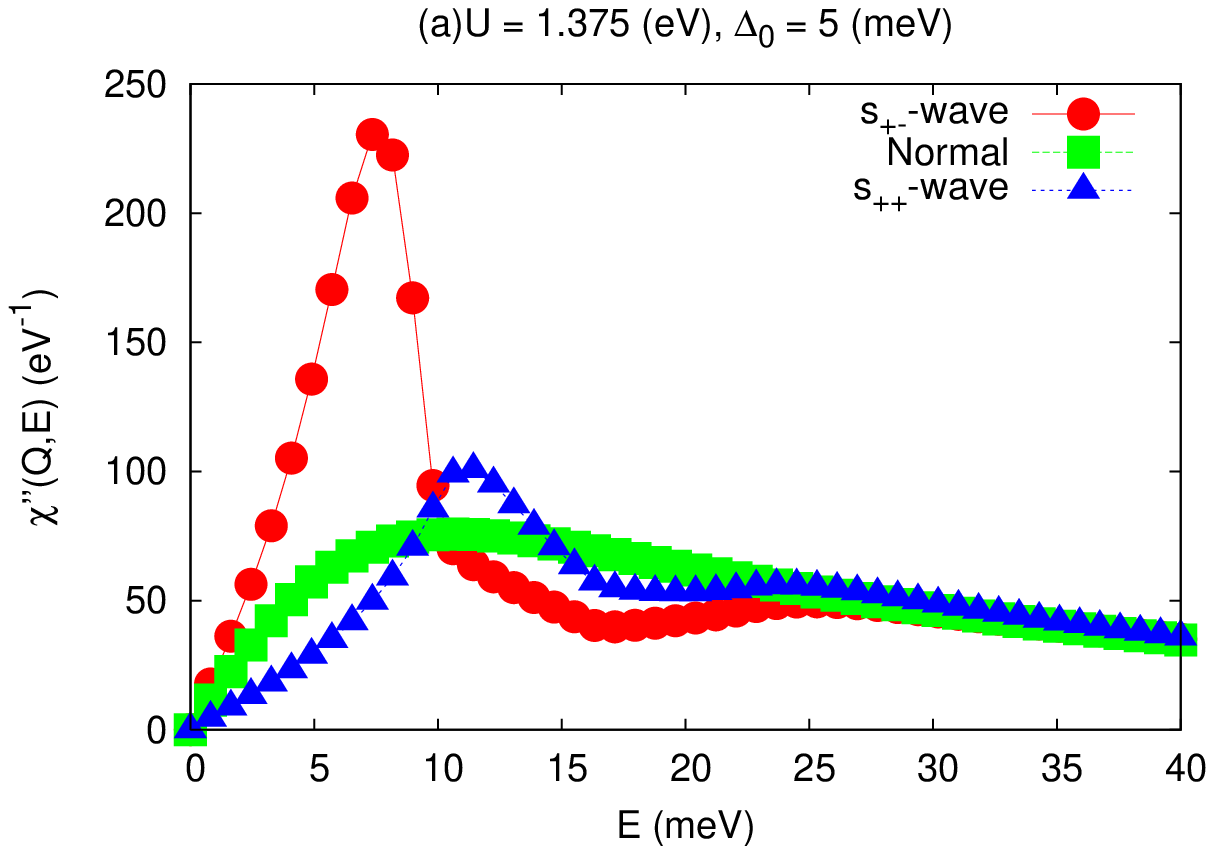}} \\
      \resizebox{65mm}{!}{\includegraphics{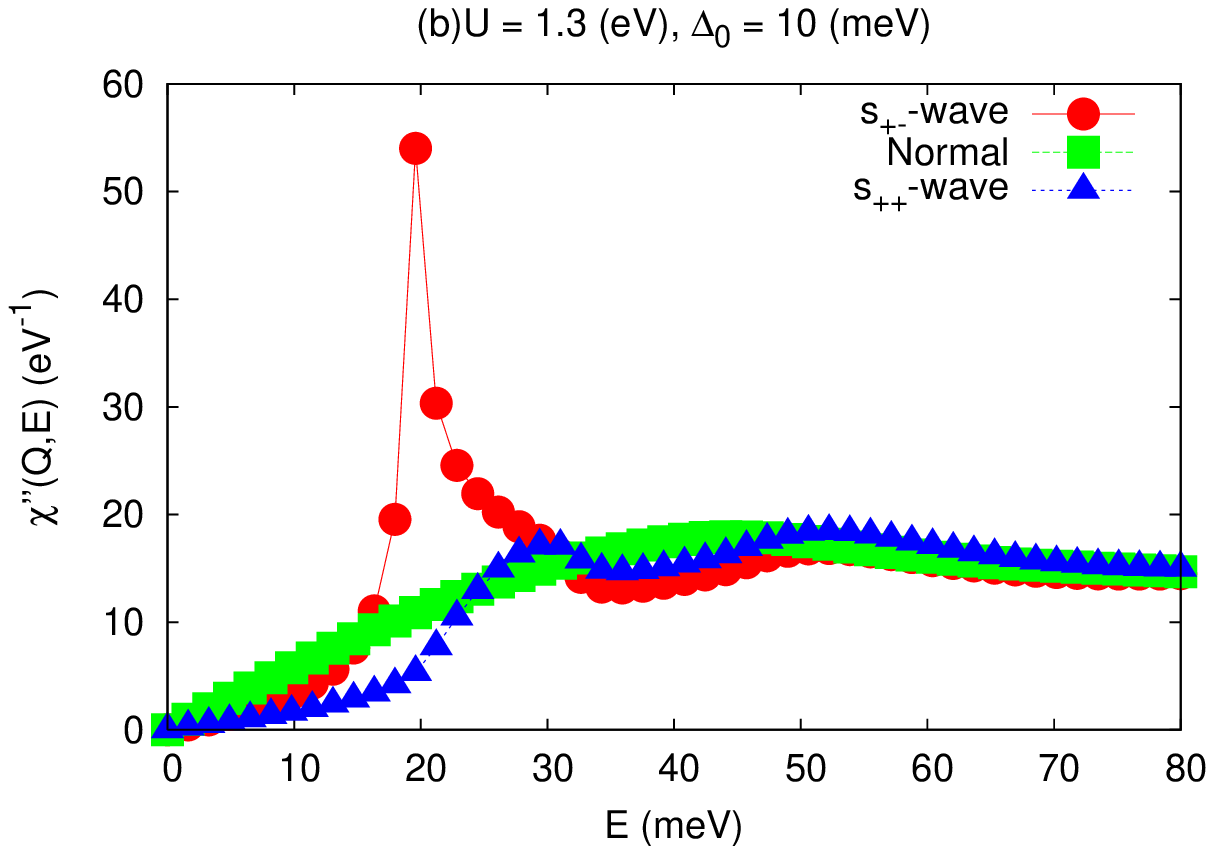}} % &
   %   \resizebox{30mm}{!}{\includegraphics{Qmap100930_qzpi.eps}} 
    \end{tabular}
\caption{\label{fig:10}
(Color online) 
Energy dependence of $\chi''(\Vec{Q},E)$ 
at $\Vec{Q} = (\pi,\pi/16)$ with the quasiparticle damping of 
$\gamma_{0} = 10$(meV). $s_\pm$ (circle), $s_{++}$ (triangle), 
and normal(square) states. We take 
(a)$\Delta_{0} = 5$(meV) and $U = 1.375$(eV), 
and (b)$\Delta_{0} = 10$(meV) and $U = 1.3$(eV). 
}
  \end{center}
\end{figure}
%%%%%%%%%
%\end{widetext}

We apply multi-orbital RPA (Random phase approximation) to the five orbital model of
LaFeAsO obtained in the unfolded Brillouin zone\cite{KurokiPRL}, 
where the $x$- and $y$- axes are taken in the Fe-Fe bond direction. 
The orbital-dependent spin 
susceptibility $\chi(\Vec{Q}, E)$ is given as 
\begin{align}
%\chi^{aa}_{cc}(\Vec{Q},E) &= \sum_{a,c}\left[ (\hat{1} - \hat{\chi}_{0}(\Vec{Q},E)   \hat{U}_{s} )^{-1} \hat{\chi}_{0}(\Vec{Q},E)  \right]^{aa}_{cc},
\chi(\Vec{Q},E) &= \sum_{a,c}\left[ (\hat{1} - \hat{\chi}_{0}(\Vec{Q},E)   \hat{U}_{s} )^{-1} \hat{\chi}_{0}(\Vec{Q},E)  \right]^{aa}_{cc},
\end{align}
where the interaction coefficients 
$U_{s}^{abcd} = U$, $U'$, $J$, and $J'$ for 
$a = b = c = d$, $a = c \neq b = d$, 
$a = b \neq c = d$, and $a = d \neq b = c$, respectively. 
Throughout this paper, we put $J = J' = 0.15$(eV), $U' = U - 2J$
 and fix the chemical potential at $\mu = 10.93$(eV) 
 , which corresponds to an optimally doped regime of $n\simeq 6.1 $ 
($10\%$ doping). 
$\hat{\chi}_{0}(\Vec{Q},E)$ is the bare spin susceptibility expressed as 
%-------------------2011/4/21 nagai --------start
\begin{align}
\left[ \chi_{0}(\Vec{Q},E) \right]_{cd}^{ab} &= - \sum_{k} \sum_{\nu \nu'}  \left[
M_{abcd}^{\nu \nu' G}(\Vec{k},\Vec{k}+ \Vec{Q}) \chi_{0 G}^{\nu \mu}(\Vec{k},\Vec{k}+\Vec{Q},E) \right. \nonumber \\
& \left. + M_{abcd}^{\nu \nu' F} (\Vec{k},\Vec{k}+ \Vec{Q}) \chi_{0 F}^{\nu \nu'}(\Vec{k},\Vec{k}+\Vec{Q},E) \right] .
\end{align}
Here, $\chi_{0 G(F)}^{\nu \nu'}(\Vec{k},\Vec{k}+\Vec{Q},E)$ denotes the normal (anomalous) part of the band-dependent BCS spin susceptibility 
written as 
\begin{align}
\chi_{0 G}^{\nu \nu'}(\Vec{k},\Vec{k}+\Vec{Q},E) &= \frac{|v_{\Vec{k}}^{\nu}|^{2}|u_{\Vec{k}+ \Vec{Q}}^{\nu'}|^{2}}{E + i \eta- E_{\Vec{k} + \Vec{Q}}^{\nu'} - E_{\Vec{k}}^{\nu}  } \\
\chi_{0 F}^{\nu \nu'}(\Vec{k},\Vec{k}+\Vec{Q},E) &= - \frac{
u_{\Vec{k}}^{\nu \ast} v_{\Vec{k}}^{\nu} u_{\Vec{k}+\Vec{q}}^{\nu'} v_{\Vec{k}+\Vec{Q}}^{ \nu' \ast}
}{E + i \eta- E_{\Vec{k} + \Vec{Q}}^{\nu'} - E_{\Vec{k}}^{\nu}  }, 
\end{align}
%-------------------2011/4/21 nagai --------end
at zero-temperature ($E > 0 $) with 
$E_{\Vec{k}}^{\nu} = \sqrt{\epsilon_{k}^{\nu 2} + |\Delta_{k}^{\nu}|^{2}}$, 
$|u_{\Vec{k}}^{\nu}|^{2} = (1 + \epsilon_{\Vec{k}}^{\nu}/E_{\Vec{k}}^{\nu})/2$, 
$|v_{\Vec{k}}^{\nu}|^{2} = (1 - \epsilon_{\Vec{k}}^{\nu}/E_{\Vec{k}}^{\nu})/2$, 
$u_{\Vec{k}}^{\nu} v_{\Vec{k}}^{\nu \ast} = \Delta_{k}^{\nu}/(2 E_{\Vec{k}}^{\nu})$, and 
$\epsilon_{\Vec{k}}^{\nu}$ is the $\nu$-th band energy measured relative to the Fermi energy. 
$M_{abcd}^{\nu \nu' G}(\Vec{k},\Vec{k}+ \Vec{Q}) $ ($M_{abcd}^{\nu \nu' F}(\Vec{k},\Vec{k}+ \Vec{Q}) $) 
is given by 
\begin{align}
M_{abcd}^{\nu \nu' G}(\Vec{k},\Vec{k}+ \Vec{Q}) &= U_{a \nu}^{\ast}(\Vec{k}) U_{b \nu'}(\Vec{k} + \Vec{Q}) U_{c \nu'}^{\ast}(\Vec{k} + \Vec{Q}) U_{d \nu}(\Vec{k}), \\
M_{abcd}^{\nu \nu' F}(\Vec{k},\Vec{k}+ \Vec{Q}) &= U_{a \nu}^{\ast}(\Vec{k}) U_{b \nu'}(\Vec{k} + \Vec{Q}) U_{c \nu}(\Vec{k}) U_{d \nu'}^{\ast}(\Vec{k}+ \Vec{Q}), 
\end{align}
with the unitary matrix $\check{U}(\Vec{k})$ which diagonalizes the 
Hamiltonian in the orbital basis. 
Here, we introduce the band-index $\nu$ whose energy $\epsilon_{\nu}$ 
satisfies the relation $\epsilon^{\nu} > \epsilon^{\nu'}$ $(\nu > \nu')$.
There are the hole Fermi surfaces on the 2nd and 3rd bands around 
$(k_{x},k_{y}) = (0,0)$ and the electron Fermi surfaces on the 4th band 
around $(\pi,0)$ and $(\pi,0)$.
For the '$s_{++}$-wave', we take  
$\Delta^{2} =  \Delta^{3 } = \Delta^{4}= \Delta_{0}$
 and for '$s_{\pm}$-wave' 
$\Delta^{2} =  \Delta^{3 } = -\Delta^{4}= \Delta_{0}$. 
As done in ref.\cite{Maier}, we 
introduce a Gaussian cutoff for the gap 
 $\Delta_{\Vec{k}}^{\nu} = \Delta^{\nu} \exp \{-[\epsilon^{\nu}_{\Vec{k}}/\Delta E]^{2} \}$, 
and take $\Delta E = 4 \Delta_{0}$. 

The quasiparticle damping 
is taken into account 
in the same form as in ref.\cite{Onari} :  
$\Gamma_{ll',\Vec{kq}} = {\rm max} \: \{ \gamma(E_{\Vec{k}}^{l}), \gamma(E^{l'}_{\Vec{k}+\Vec{q}}) \}$ with 
$\gamma(\epsilon) = a(\epsilon) \gamma_{0}$ where $\gamma(\epsilon) = \eta$ for $|\epsilon| < 3 \Delta_{0}$, $\gamma(\epsilon) = \gamma_{0}$ for $|\epsilon| > 4 \Delta_{0}$, and linear interpolation for $3 \Delta_{0} < |\epsilon| < 4 \Delta_{0}$.\cite{Onari}  
In the present study, the smearing factor 
is taken to be $\eta = 0.5$(meV) in order to consider 
small values ($\sim 10$(meV)) of $\gamma_0$.
To cope with the realistic magnitude of the  superconducting gap, 
we take $8192 \times 8192$ $\Vec{k}$-point meshes throughout the paper.
We have confirmed that the results do not change 
if we take 16384 $\times$ 16384 $\Vec{k}$-point meshes.  
 
We first calculate $\chi''(\Vec{Q},E)={\rm Im} \chi(\Vec{Q}, E)$ 
at the nesting vector of the electron and hole Fermi surfaces 
$\Vec{Q} = (\pi,\pi/16)$ for $s_{\pm}$ and $s_{++}$ states.
We adopt the values  $\Delta_{0} = 5$(meV) in 
(a) and $\Delta_{0} = 10$(meV) in (b) for the superconducting gap.
We take $\gamma_{0} = 10$(meV) because 
$\gamma_0$ is estimated to be of the same order as $\Delta_0$\cite{Onari}. 
The values of $U$ are determined so that the normal state 
susceptibility is broadly maximized around several times of $\Delta_0$
as observed experimentally.
As shown in Fig.~\ref{fig:10}, the resonance peak develops above the
normal state value for the $s_{\pm}$-wave state, which  
is qualitatively consistent with the 
previous studies\cite{Maier2,Maier,KorshunovPRB2008}.
In contrast to the case with larger quasiparticle damping and
superconducting gap\cite{Onari}, 
the spectral weight at the peak is only about three times larger 
compared to the normal state value, which 
can be considered as consistent with the experimental results
~\cite{Inosov,Christianson,Qiu,Zhao,Ishikado}.
We note here that not only the small $\Delta_0$ and $\gamma_0$, 
but also the Gaussian cutoff for the superconducting gap 
is also effective in 
reducing the spectral weight in the superconducting state\cite{comment3D}.
On the other hand,  there is only a small enhancement 
in the case of $s_{++}$ again in contrast to the case with 
larger quasiparticle damping and superconducting gap\cite{Onari}. 
Thus, within the present formalism, as far as we adopt the $\gamma_0$ value 
estimated from experimental results\cite{Onari} and the 
realistic magnitude of the superconducting gap,  
the $s_{\pm}$ state is more consistent with the experiments. 

%%%%%%%%%

\begin{figure}[t]
  \begin{center}
    \begin{tabular}{p{65mm}}%  p{28mm}}
      \resizebox{65mm}{!}{\includegraphics{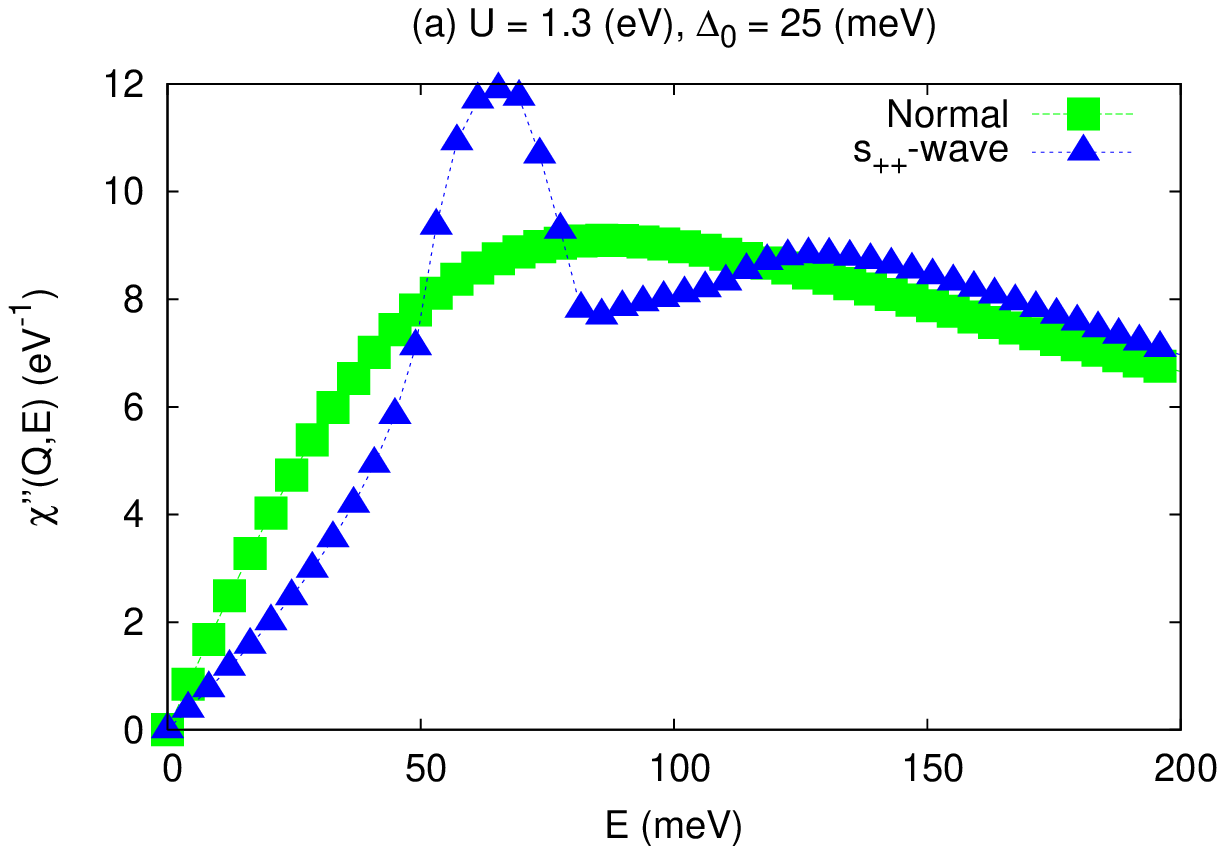}} \\
      \resizebox{65mm}{!}{\includegraphics{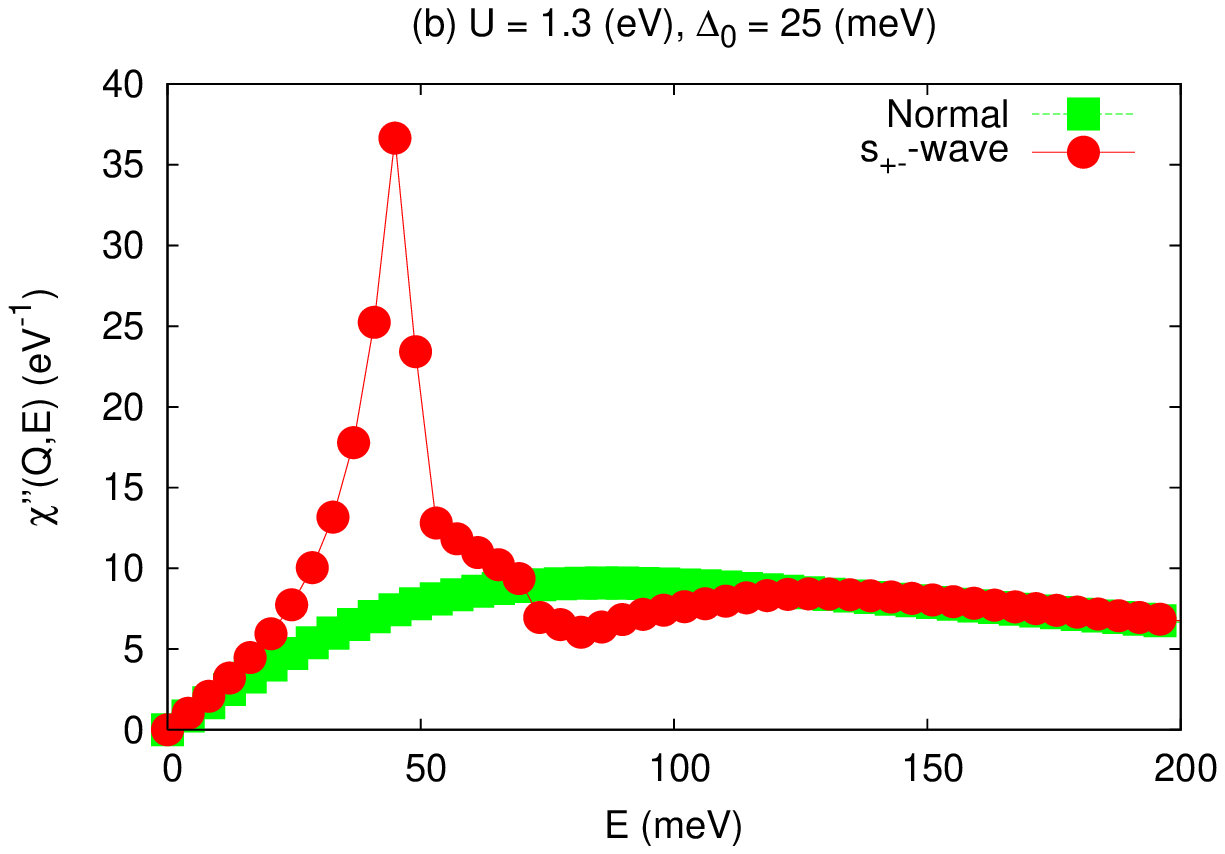}} % &
   %   \resizebox{30mm}{!}{\includegraphics{Qmap100930_qzpi.eps}} 
    \end{tabular}
\caption{\label{fig:50}
(Color online) 
Energy dependence of $\chi''(\Vec{Q},E)$ 
at $\Vec{Q} = (\pi,\pi/16)$ 
in the case of (a) $s_{++}$ and (b) $s_{\pm}$ with 
$\Delta_{0} = 25$(meV), the quasiparticle damping $\gamma_{0} =
 50$(meV), 
% various $\Vec{Q}$-vectors with $\Delta_{0} = 10$(meV). 
and $U = 1.3$(eV).
% We put $\gamma(\epsilon) = a(\epsilon) \gamma_{0}$, $\gamma_{0} = \Delta_{0}$ and $U = 1.3$(eV).
}
  \end{center}
\end{figure}

To see how the peak like structure develops 
in the $s_{++}$ state for stronger 
quasiparticle damping, we now calculate $\chi''(\Vec{Q}=(\pi,\pi/16),E)$ 
with $\gamma_{0} = 2 \Delta_0=50$(meV) as shown in Fig.~\ref{fig:50}(a).
We see that the peak structure indeed  develops for such a large 
$\gamma_{0}$ and $\Delta_0$,  consistent with ref.\cite{Onari}. 
The origin of the enhancement above the normal state values  
is mainly due to the the strong suppression of
$\chi''$ in the normal state by the quasiparticle damping.

For such a large $\gamma_0$ and $\Delta_0$, 
the enhancement ratio of the 
resonance peak in the $s_{\pm}$ state against the normal state value  
becomes somewhat larger as shown in Fig.~\ref{fig:50}(b). 
In fact, for even larger values of $\Delta_0$ and $\gamma_0$, 
the resonance peak in the $s_{\pm}$ state becomes much larger compared to the 
experiments, as was shown in ref.\cite{Onari}.
Thus, as we increase the values  of $\gamma_0$ and $\Delta_0$ 
by an order of magnitude larger than the experimentally evaluated 
values, 
there is a tendency that the $s_{++}$ state becomes more consistent with the 
experiments than the $s_{\pm}$ state.

%%%

%%%%%%%%%%
%%%
\begin{figure}[bt]
  \begin{center}
    \begin{tabular}{p{65mm}}%  p{28mm}}
      \resizebox{65mm}{!}{\includegraphics{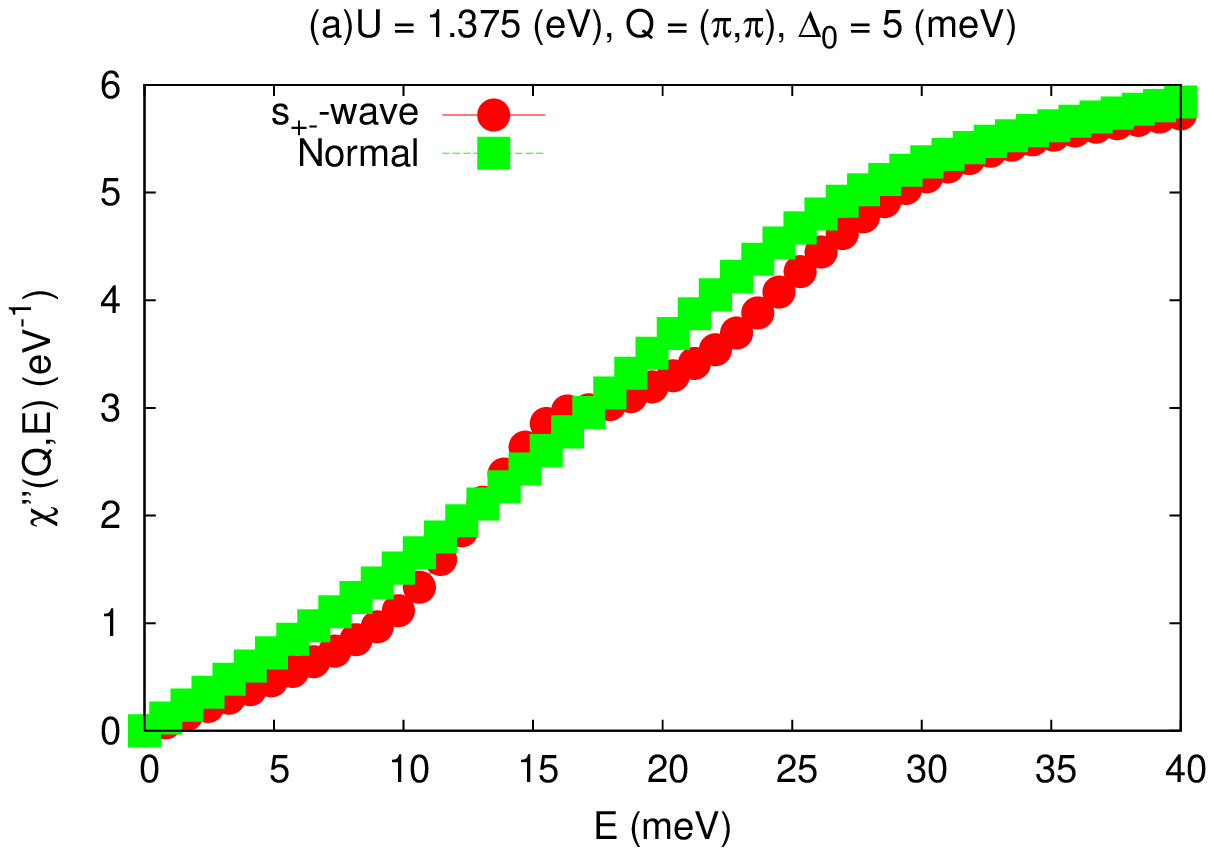}} \\
      \resizebox{65mm}{!}{\includegraphics{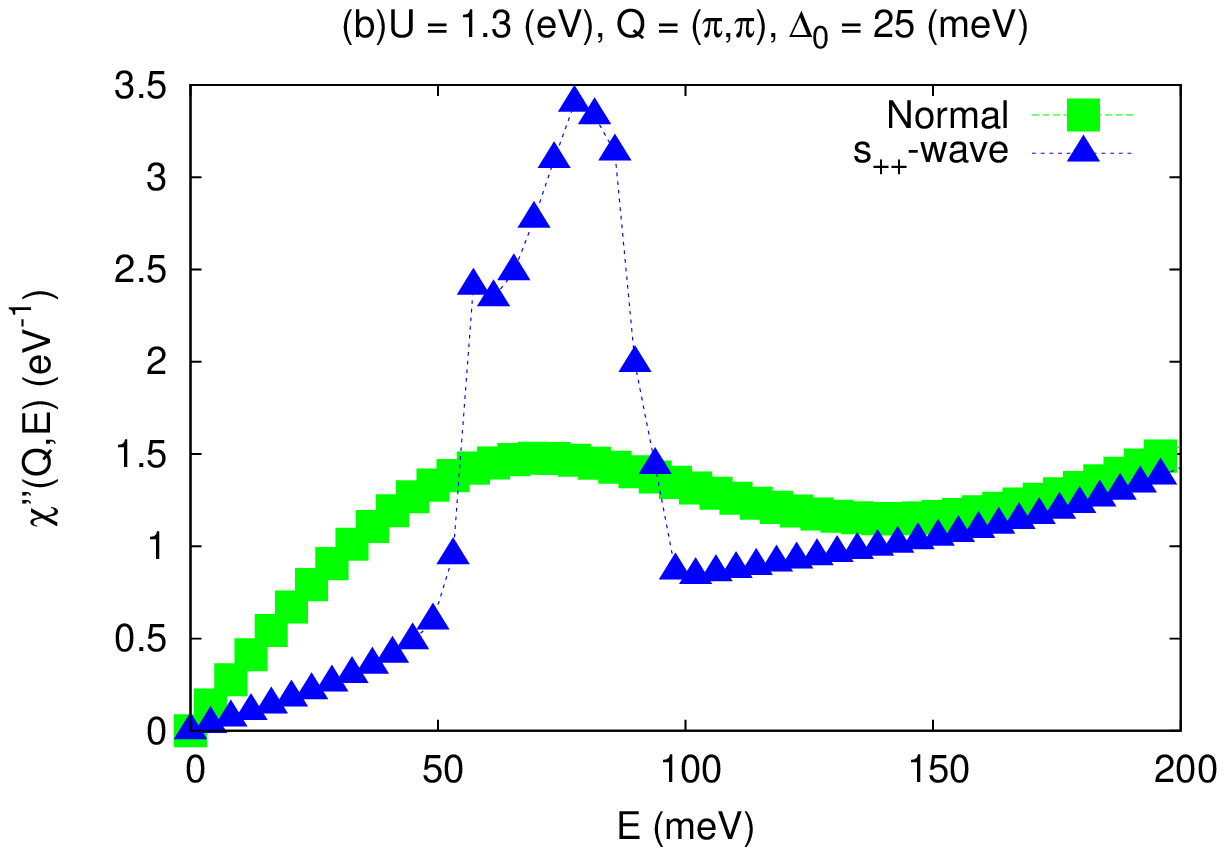}} % &
   %   \resizebox{30mm}{!}{\includegraphics{Qmap100930_qzpi.eps}} 
    \end{tabular}
\caption{\label{fig:5pi}
(Color online) 
Energy dependence of $\chi''(\Vec{Q},E)$ 
at $\Vec{Q} = (\pi,\pi)$ for parameter values where 
(a)$s_{\pm}$ and (b)$s_{++}$ states are consistent with the 
experiments at the wave vector $\simeq (\pi,0)$.
(a)$\Delta_{0} = 5$(meV), $U = 1.375$(eV) and $\gamma_{0} = 10$(meV), 
and (b)$\Delta_{0} = 25$(meV), $U = 1.3$(eV) and $\gamma_{0} = 50$(meV).
%various gap amplitudes $\Delta_{0} = 30,  50$(meV), respectively. 
%We consider the energy-dependent quasiparticle damping. We put $\gamma(\epsilon) = a(\epsilon) \gamma_{0}$, $\gamma_{0} = \Delta_{0}$ and $U = 1.3$(eV).
}
  \end{center}
\end{figure}

So far, our results for the wave vector $\Vec{Q} = (\pi,\pi/16)$
show that the $s_{\pm}$ state is consistent
with the experiments when small values of $\gamma_0$ and 
$\Delta_0$ are adopted, while $s_{++}$ can explain the
experiments for large $\gamma_0$ and $\Delta_0$. 
The former parameter values are realistic as far as the 
experimental evaluations are concerned, but there 
might remain a possibility that the 
{\it effective} values of $\gamma_0$ and/or $\Delta_0$ 
that should be adopted in the present formalism 
turn out to be larger\cite{comment}.
In that sense, the possibility of $s_{++}$ state may not be 
ruled out. 
Here we propose a way to distinguish whether the peak like structure 
at $\sim (\pi,0)$ 
originates from the resonance effect in the $s_{\pm}$ state or the 
strong quasiparticle damping effect in the $s_{++}$ state.
We calculate $\chi''(\Vec{Q},E)$ at $\Vec{Q} = (\pi,\pi)$ . 
$\Vec{Q} = (\pi,\pi)$ is the translation vector 
for which the two electron Fermi
surfaces around $(\pi,0)$ is nearly superposed to that around 
$(0,\pi)$. (It should be noted however that this wave vector is 
not the nesting vector that induces spin fluctuations since 
this translation superposes an electron Fermi surface on another 
electron Fermi surface.)
We consider the two cases where the experimental results can be explained 
at $\Vec{Q} =(\pi,\pi/16)$ : 
(a) $s_{\pm}$-wave with $\gamma=2\Delta_{0} =10$(meV) and 
(b) $s_{++}$-wave with $\gamma=2\Delta_{0} = 50$(meV). 
%--------------2011/5/19 nagai-------------start
%---- kuroki 5/19 ------ 
Here we consider only the case of $s_{\pm}$ $(s_{++})$ for small (large) 
$\gamma$ and $\Delta_0$ because this is the combination that 
is consistent with the experiment at $\simeq (\pi,0)$, but 
we note here that actually, for each fixed parameter set, 
the calculation results at $(\pi,\pi)$ 
are barely affected by whether the gap is $s_{\pm}$ 
or $s_{++}$ because in both cases 
the superconducting gap {\it does not} change its sign between 
the Fermi surfaces separated by $(\pi,\pi)$.
%--------------2011/5/19 nagai-------------end
As shown in Fig.~\ref{fig:5pi}(a), no enhancement above the normal 
state values is found in the former case because the  
quasiparticle damping is small. 
On the other hand, in the latter case, 
we find a peak structure developing above the 
normal state values {\it even} at $(\pi,\pi)$  because 
the effect of the quasiparticle damping is strong. 
Therefore, by investigating the spin fluctuations at $\simeq (\pi,\pi)$ using 
neutron scattering experiments, one can distinguish the origin of the 
peak like structure at $\sim (\pi,0)$ ; if an enhancement above the
normal state values is found in the superconducting state 
at $(\pi,\pi)$ as well as at $(\pi,0)$, its origin may well be due to 
the quasiparticle damping, while if an enhancement is observed at 
$(\pi,0)$ but not at $(\pi,\pi)$, the quasiparticle damping is not
so strong and the origin is the resonance in the $s_{\pm}$ state.

In conclusion, 
we have calculated the spin susceptibility in the 
$s_{\pm}$ and $s_{++}$ superconducting states of the 
iron pnictides by applying 
multi-orbital RPA to the 
effective five-band model and considering the quasiparticle damping.
We have found that as far as we adopt the values of the 
quasiparticle damping and the superconducting gap evaluated from 
experiments, the enhancement above the normal 
state values at $\sim (\pi,0)$ observed in the neutron scattering 
experiments is more consistent with the resonance peak in the 
$s_{\pm}$ state. On the other hand, for large magnitude of the 
quasiparticle damping and the superconducting gap, 
the enhancement in the spin susceptibility of the $s_{++}$ state  
over the normal state values 
can explain the experimental results\cite{Onari}, while the 
resonance peak in the $s_{\pm}$ state becomes large compared to 
experimental observations.
In case there is a possibility that the quasiparticle damping 
and the superconducting gap that should be adopted in the 
present formalism is effectively larger than the experimentally 
evaluated values, we propose 
to investigate in the neutron scattering experiments the wave 
vector around $(\pi,\pi)$ in the unfolded Brillouin zone 
to see if there is an enhancement like those found 
at $(\pi,0)$ in the superconducting state.

We thank M. Machida, N. Nakai, Y. Ota, S. Shamoto, M. Ishikado, 
S. Onari, H. Kontani, D.J. Scalapino, and H. Aoki 
for helpful discussions and comments.
The calculations have been performed using the supercomputing 
system PRIMERGY BX900 at the Japan Atomic Energy Agency.


\begin{thebibliography}{99}
\bibitem{Kamihara}
Y. Kamihara, T. Watanabe, M. Hirano, and H. Hosono, 
J. Am. Chem. Soc. {bf 130}, 3296 (2008).
\bibitem{BangPRB2008}
Y. Bang, and H.-Y. Choi, 
Phys. Rev. B, {\bf 78}, 134523 (2008).
\bibitem{SeoPRL2008}
K. Seo, B. A. Bernevig, and J. Hu, 
Phys. Rev. Lett. {\bf 101}, 206404 (2008). 
\bibitem{ParishPRB2008}
M. M. Parish, J. Hu,   and B. A. Bernevig, 
Phys. Rev. B {\bf 78}, 144514 (2008).
\bibitem{KorshunovPRB2008}
M. M. Korshunov and I. Eremin, Phys. Rev. B {\bf 78}, 140509(R) (2008). 
\bibitem{Mazin}
I. I. Mazin, D. J. Singh, M. D. Johannes, and M. H. Du,
Phys. Rev. Lett. {\bf 101}, 057003 (2008).
\bibitem{KurokiPRL}
K. Kuroki, R. Arita, H. Usui, Y. Tanaka, H. Kontani, and H. Aoki,
Phys. Rev. Lett. {\bf 101}, 087004 (2008).
\bibitem{Hashimoto}
K. Hashimoto, T. Shibauchi, T. Kato, K. Ikada, R. Okazaki, H. Shishido,
M. Ishikado, H. Kito, A. Iyo, H. Eisaki, S. Shamoto, and Y. Matsuda, 
Phys. Rev. Lett. {\bf 102}, 017002 (2009).
%arXiv:0806.3149. %Pr BCS
\bibitem{LuetkensPRL2008}
H. Luetkens, H.-H. Klauss, R. Khasanov, A. Amato, R. Klingeler, I. Hellmann, N. Leps, A. Kondrat, C. Hess, 
A. K\"{o}hler, G. Behr, J. Werner, and B. B\"{u}chner, 
Phys. Rev. Lett. {\bf 101}, 097009 (2008). 
\bibitem{MaronePRB2009}
L. Malone, J.~D. Fletcher, A. Serafin, A. Carrington, N.D. Zhigadlo, Z. Bukowski, S. Katrych, and J. Karpinski, 
Phys. Rev. B {\bf 79}, 140501(R) (2009). 
\bibitem{DingEPL}%122ARPES
H. Ding, P. Richard, K. Nakayama, T. Sugawara, T. Arakane, Y. Sekiba, A. Takayama, S. Souma, T. Sato, T. Takahashi, Z. Wang, X. Dai, Z. Fang, G. F. Chen, J. L. Luo, and N. L. Wang, 
EuroPhys. Lett. {\bf 83}, 47001 (2008). 
\bibitem{Nakamura}
K. Nakamura, T. Sato, P. Richard, Y.-M. Xu, Y. Sekiba, S. Souma, G. F. Chen, 
J. L. Luo, N. L. Wang, H. Ding, and T. Takahashi, 
Europhys. Lett. {\bf 85} 67002 (2009).
\bibitem{Evtushinsky}%122ARPES scGap
D. V. Evtushinsky, D. S. Inosov, V. B. Zabolotnyy, A.Koitzsch, M. Knupfer, B. Buchner, M. S. Viazovska, G. L. Sun, V. Hinkov, A. V. Boris, C. T. Lin, B. Keimer, A. Varykhalov, A. A. Kordyuk, and S. V. Borisenko, 
Phys. Rev. B {\bf 79}, 054517 (2009). 
\bibitem{Nakai}
Y. Nakai, K. Ishida, Y. Kamihara, M. Hirano, and H. Hosono,
J. Phys. Soc. Japan, {\bf 77} 073701 (2008). 
\bibitem{Grafe}
H.-J. Grafe, D. Paar, G. Lang, N.~J. Curro, G. Behr, J. Werner, J.
  Hamann-Borrero, C. Hess, N. Leps, R. Klingeler, and B. B\"{u}chner, 
 Phys. Rev. Lett. {\bf 101}, 047003 (2008).
  \bibitem{MukudaJPSJ08}
H. Mukuda, N. Terasaki, H. Kinouchi, M. Yashima, Y. Kitaoka, S. Suzuki, S.
  Miyasaka, S. Tajima, K. Miyazawa, P. Shirage, H. Kito, H. Eisaki, and A. Iyo, 
J. Phys. Soc. Jpn. {\bf 77}, 093704 (2008).
    \bibitem{TerasakiFeNMR}
N. Terasaki, H. Mukuda, M. Yashima, Y. Kitaoka, K. Miyazawa, P.~M. Shirage, H.
  Kito, H. Eisaki, and A. Iyo, 
J. Phys. Soc. Jpn. {\bf 78}, 013701 (2009).
\bibitem{IBM} C.-T. Chen, 
C. C. Tsuei, M. B. Ketchen, Z.-A. Ren, and Z. X. Zhao, Nature Physics {\bf 6}, 260 (2010).
\bibitem{HanaguriFe}
T. Hanaguri, S. Nittaka, K. Kuroki, and H. Takagi, Science {\bf 328}, 474 (2010). 
\bibitem{Guo}
Y. F. Guo, Y. G. Shi, S. Yu, A. A. Belik, Y. Matsushita, M. Tanaka, Y. Katsuya, K. Kobayashi, I. Nowik, I. Felner, V. P. S. Awana, K. Yamaura, 
and E. Takayama-Muromachi, 
Phys. Rev. B {\bf 82}  054506 (2010). 
  \bibitem{Sato}
M. Sato, Y. Kobayashi, S. C. Lee, H. Takahashi, E. Satomi, and Y. Miura, 
J. Phys. Soc. Jpn. {\bf 79} (2010) 014710. 
  \bibitem{Li}
Y. K. Li, X. Lin, Q. Tao, C. Wang, T. Zhou, L. J. Li, Q. B. Wang, M. He, G. H. Cao, and Z. A. Xu, 
New J. Phys. {\bf 11}  053008 (2009).
\bibitem{OnariKontani}
S. Onari and H. Kontani, Phys. Rev. Lett. {\bf 103} (2009) 177001.
\bibitem{NakamuraArita}
K. Nakamura, R. Arita, and H. Ikeda, Phys. Rev. B {\bf 83}, 144512 (2011).
%arXiv:1010.0533 (unpublished).
\bibitem{Kontani}
H. Kontani and S. Onari, 
Phys. Rev. Lett. {\bf 104}, 157001 (2010). 
\bibitem{Ono}
Y. Yanagi, Y. Yamakawa, N. Adachi, and Y. Ono, 
J. Phys. Soc. Jpn. {\bf 79},123707 (2010).
%arXiv:1010.0129.
\bibitem{Saito}
T. Saito, S. Onari, and H. Kontani, 
Phys. Rev. B {\bf 82} 144510 (2010). 
\bibitem{Maier2} T.A. Maier and D. J. Scalapino, Phys. Rev. B {\bf 78}, 020514(R) (2008). 
\bibitem{Maier}
T. A. Maier, S. Graser, D. J. Scalapino, and P. Hirschfeld, 
Phys. Rev. B {\bf 79}, 134520 (2009). 


\bibitem{Christianson}
A. D. Christianson, E. A. Goremychkin, R. Osborn, S. Rosenkranz, M. D. Lumsden, C. D. Malliakas, I. S. Todorov, H. Claus, D. Y. Chung, M. G. Kanatzidis, R. I. Bewley  and  T. Guidi, 
Nature {\bf 456} 930 (2008).
\bibitem{Qiu}
Y. Qiu, W. Bao, Y. Zhao, C. Broholm, V. Stanev, Z. Tesanovic, Y. C. Gasparovic, S. Chang, J. Hu, B. Qian, M. Fang, and Z. Mao, 
Phys. Rev. Lett. {\bf 103} 067008 (2009).   
\bibitem{Inosov}
D. S. Inosov, J. T. Park, P. Bourges, D. L. Sun, Y. Sidis, A. Schneidewind, K. Hradil, D. Haug, C. T. Lin, B. Keimer and V. Hinkov, 
Nature Phys. {\bf 6} 178 (2010). 
\bibitem{Zhao}
J. Zhao, L.-P. Regnault, C. Zhang, M. Wang, Z. Li, F. Zhou, Z. Zhao, C. Fang, J. Hu and P. Dai, Phys. Rev. B {\bf 81}, 180505(R) (2010). 
\bibitem{Ishikado}
M. Ishikado, Y. Nagai, K. Kodama, R. Kajimoto, M. Nakamura, Y. Inamura, S. Wakimoto, H. Nakamura, M. Machida, K. Suzuki, H. Usui, K. Kuroki, A. Iyo, H. Eisaki, M. Arai, S. Shamoto, 
arXiv:1011.3191 (unpublished). 
\bibitem{Onari}
S. Onari, H. Kontani, and M. Sato, 
Phys. Rev. B {\bf 81} 060504(R) (2010).
%  --- kuroki 2011/5/18-----start
\bibitem{comment3D} The present calculation has been done for a purely 
two dimensional model of LaFeAsO. On the other hand, the 122 materials
exhibit strong three dimensionality, and 
a possibility of superconducting gap nodes or dips  on the 
strongly warped portion of the hole Fermi surface  has been 
pointed out theoretically (
S. Graser {\it et al.}, Phys. Rev. B {\bf 81}, 214503  (2010) and 
K. Suzuki {\it et al.}, J. Phys. Soc. Jpn. {\bf 80}, 103710 (2011)) 
or experimentally 
(e.g., F. Hardy, {\it et al.}, EuroPhys. Lett. {\bf 91} (2010) 47008).
In ref.~\cite{Ishikado}, we have performed a 
calculation for a three dimensional model of 122 material 
(without  considering quasi particle damping),  
where taking into account the above mentioned effect 
tends to reduce the strength of the resonance 
peak in the $s_{\pm}$ state, which gives further agreement with experiments.
%  --- kuroki 2011/5/18-----end





\bibitem{comment} For instance, the energy scale of the 
superconducting transition temperature and the superconducting gap 
in the random phase approximation 
tends to be larger compared to actual experimental values. 

%\bibitem{Lee}
%I. J. Lee, 
%S. E. Brown, W. G. Clark, M. J. Strouse, M. J. Naughton, W. Kang, and P. M. Chaikin, 
%Phys. Rev. Lett. {\bf 88}, 017004 (2002).   
%\bibitem{Suzuki1010}
%K. Suzuki, H. Usui, and K. Kuroki, arXiv:1010.3542.
%\bibitem{Miyake}
%T. Miyake, K. Nakamura, R. Arita, M. Imada, J. Phys. Soc. Jpn. {\bf 79} 044705 (2010). 


\end{thebibliography}
\end{document}